\documentclass[aps,prb,twocolumn,amssymb,showpacs]{revtex4}

\usepackage{graphicx}
\usepackage{dcolumn}
\usepackage{bm}

\begin{document}

\title{Magnetism and Electronic Correlations in Quasi-One-Dimensional Compounds}
\author{M.~D. Coutinho-Filho}
\email{mdcf@ufpe.br}
\author{R.~R. Montenegro-Filho}
\email{rene@df.ufpe.br}
\author{E.~P. Raposo}
\email{ernesto@df.ufpe.br}
\affiliation{Laborat\'orio de F\'{\i}sica
Te\'orica e Computacional, Departamento de F\'{\i}sica,
Universidade Federal de Pernambuco, 50670-901, Recife-PE, Brazil}
\author{C. Vitoriano}
\email{carlindovitoriano@ig.com.br}
\affiliation{Universidade Federal Rural de Pernambuco, 
Unidade Acad\^emica de Garanhuns, 55296-190, Garanhuns-PE, Brazil}
\author{M. H. Oliveira}
\email{mario@lftc.ufpe.br}
\affiliation{Universidade Federal Rural de Pernambuco, 
Unidade Acad\^emica de Serra Talhada, 56900-000, Serra Talhada-PE, Brazil}
\date{\today}
\begin{abstract}
In this contribution on the celebration of the 80th birthday anniversary 
of Prof. Ricardo Ferreira, we present a brief survey on 
the magnetism of quasi-one-dimensional 
compounds. This has been a research area of intense activity 
particularly since the first experimental announcements 
of magnetism in organic and organometallic polymers in the mid 80's. 
We review experimental and theoretical 
achievements on the field, featuring chain systems of correlated electrons 
in a special $AB_2$ unit cell 
structure present in inorganic and organic compounds. 
%
%
\end{abstract}
\maketitle
\section{Introduction}
The magnetism of organic \cite{r1,r10,yo1} and organometallic \cite{r22} polymers has been a challenging topical field since 
its first experimental announcements. This rapidly 
growing and interdisciplinary research area also includes 
inorganic compounds \cite{r6}, with ferro- and ferrimagnetic long-range order (see Sec.~II below). 
In this work we briefly review some attempts to describe the ground state and 
the low-temperature thermodynamics of these compounds.  
In particular, we report on analytical and numerical results 
on polymeric chains of correlated electrons in special unit cell topologies shown in Fig.~\ref{f1}. 
\section{Quasi-One-Dimensional Magnetic Compounds: A Brief Review}
Despite many years of experimental and theoretical efforts, 
the complete understanding and precise characterization of  
magnetism and electronic correlations in quasi-one-dimensional (quasi-$1d$) compounds still offer great 
scientific challenges and technical difficulties~\cite{yo1}. 
Regarding, for instance, organic magnetic polymers, 
it is known~\cite{yo1} that their magnetic properties are 
ascribed to the correlated p-electrons of light 
elements, such as C, O, and N, in contrast to the magnetism found in 
transition and rare-earth metals due to partially filled d or f orbitals.
For this reason it took many years of efforts on the synthesis and 
characterization of a great variety of compounds before the announcement 
of bulk ferromagnetism (FM) in an organic polymer~\cite{r1}.
In Fig.~\ref{f1}(b) we sketch this material made of polyacetilene-based 
radicals, $R^{\ast}$, containing unpaired residual electrons, i.e., 
{\it poly}-BIPO or poly[1,4-bis(2,2,6,6-tetramethyl-4-piperidyl-1-oxyl)-butadiyne].
However, this compound presented several problems due to 
its insolubility and poor reproducibility both in the 
preparation and in the magnetic results~\cite{yo1}. 
Later, Nishide and collaborators~\cite{nishide} have successfully 
synthesized polyphenyacetylenes with various types of radical groups. 
These polymers exhibit similar band structure schemes~\cite{r7} comprising 
filled bonding molecular-orbital bands, empty antibonding bands, 
and narrow half-filled nonbonding bands, usually just one at the center
of the band. A net magnetic moment may appear either because 
the number of itinerant antiferromagnetically (AF)-correlated $\pi$ 
electrons per unit cell is odd and/or due to the presence of localized 
electrons~\cite{r1,r7}.
\newline {\mbox \quad}A seminal work has also been 
performed by Takahashi and collaborators in order to extensively characterize the long-range 
macroscopic FM behavior found in the organic compound   
$p$-nitrophenyl nitroxyl nitroxide radical ($p$-NPNN, in the 
$\gamma$ and $\beta$ phases)~\cite{r10,r11}.
Actually, the excellent fitting of the low-temperature ($T$) experimental data 
is consistent with predictions from 
the thermodynamic Bethe-{\it ansatz} solution of the  
$1d$-quantum FM 
Heisenberg model~\cite{r11}: susceptibility $\chi \sim T^{-2}$ 
and specific heat $C \sim T^{1/2}$, as $T \rightarrow 0$.
\newline {\mbox \quad}Other organic magnets have been 
synthesized, such as polyradicals derived from poly(1,3-phenylenemethylene) 
and polyphenylenevinylene-based radicals~\cite{yo1}. In these 
cases the polymer structure is made of benzene rings linked by 
divalent carbon atoms or including pendant radicals with oxygen atoms carrying 
an uncompensated electron~\cite{yo1}. 
Another family of organic magnetic polymers is that of
the doped poly($m$-aniline) compounds, 
in which the carbon atoms responsible for the links between 
the benzene rings are substituted by ionized nitrogen with a H-bond 
or a radical plus a charge acceptor~\cite{r8}. 
On the other hand, doped polypyrrole compounds also 
exhibit\cite{r9} interesting magnetic properties and Drude metallic response 
as well. 
\newline {\mbox \quad}A distinct class of magnetic polymers combines metal ions 
with organic complexes,  
displaying a rich variety of magnetic behaviors, such as ferro- and ferrimagnetism, 
AF and canted AF and spin glass phase~\cite{r21,yo15}. In fact, the first 
experimental observation of a magnet with spin residing in a p-orbital was
performed in the compound 
[Fe(C$_5$Me$_5$)$_2$]$^+$[TCNE]$^-$ 
(TCNE $=$ tetracyanoethylene)~\cite{r22}. 
Some homometallic ferrimagnets with chain structure~\cite{yo19,yo20} involve 
the compounds~\cite{yo21,yo22} $M_2$(EDTA)(H$_2$O)$_4 \cdot$2H$_2$O 
($M = $ Ni, Co; EDTA $=$ ethylenediamminetetraacetate $=$ 
C$_{10}$N$_2$O$_8$) and $M(R$-py)$_2$(N$_3)_2$ ($M = $Cu, Mn; 
$R$-py = pyridinic ligand $=$ C$_5$H$_4$N-$R$ with $R = $ Cl, CH$_3$, 
etc.)~\cite{yo23,yo24,yo25,yo26}. 
Regarding bimetallic chain materials, the compound~\cite{yo2} MnCu(pbaOH)(H$_2$O)$_3$ 
[pbaOH $=$ 2-hydroxy-1, 3-propylenebis(oxamato) $=$ 
C$_7$H$_6$N$_2$O$_7$] has been one of the first synthesized 
which retains long-range FM or ferrimagnetic order on the scale of the crystal lattice, 
as in the case of isomorphous realizations~\cite{yo3,re29,re30}. 
Heterometallic chain structures have also been object of systematic 
study~\cite{yo4,yo5}. 
More recently, the metal-radical hybrid strategy, 
combined with fabrication of novel polyradicals~\cite{yo16}, has led to the synthesis 
of a variety of heterospin chain compounds~\cite{yo17,yo18}. 
Several of these compounds display $1d$ ferrimagnetic behavior~\cite{r26,r27} modeled
by alternating spin chains~\cite{r28}, such as, for instance, those 
with the structure shown in Fig.~\ref{f1}(c). 
\newline {\mbox \quad}Of recently growing interest we mention 
the quasi-$1d$ chains with $AB_2$ and $ABB'$ unit cell structure 
[henceforth referred to as $AB_2$ chains; see Figs.~\ref{f1}(a) and (b)]. 
Such structures are found both in inorganic and organic ferrimagnetic compounds.
Regarding inorganic materials, we cite the homometallic compounds 
with a line of trimer clusters characteristic  
of phosphates of formula A$_3$Cu$_3$(PO$_4$)$_4$, where 
A = Ca,~\cite{re23,re24,re25,re26}, Sr,~\cite{re24,re25,re26,re27}, 
and Pb~\cite{re25,re26,re28}. 
The trimers have three Cu$^{+2}$ paramagnetic ions of spin $S = 1/2$ 
AF coupled. 
Although the superexchange interaction is much weaker than the 
intratrimer coupling, it proves sufficient to turn them into bulk 
ferrimagnets. Furthermore, compounds of formula Ca$_{3-x}$Sr$_x$Cu$_3$(PO$_4$)$_4$, 
$0 \le x \le 3$~\cite{re27}, hybrid analogous to the mentioned phosphates, 
have also been synthesized in an attempt to tune the 
AF bridges between Cu sites and possibly explore how 
paramagnetic spins grow into bulk ferrimagnets. 
It is also interesting to mention the frustrated $AB_2$ 
inorganic compound~\cite{kikuchi} 
Cu$_3$(CO$_3$)$_2$(OH)$_2$, which displays low-$T$ short-range magnetic order and has 
its physical properties well described through the distorted diamond chain model \cite{RevFrus}.  
At last, we observe that the $AB_2$ structure is also present~\cite{re31} in the 
organic ferrimagnetic compound
2-[3',5'-bis($N$-$tert$-butylaminoxyl)phenyl]-4,4,5,5-tetramethyl-4,5-dihydro-1$H$-imidazol-1-oxyl 3-oxide, 
or PNNBNO, 
consisting of three $S = 1/2$ paramagnetic radicals in 
its unit cell. 
\section{Magnetic Chains with $AB_2$ Unit Cell Topology: Analytical and Numerical Studies}
\begin{figure}
\begin{center}
\includegraphics*[width=0.9\linewidth,clip]{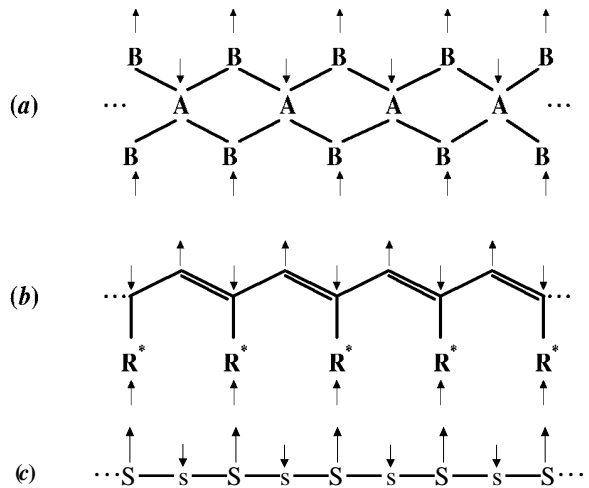}
\caption{Ferrimagnetic ground-state configurations of (a) bipartite lozenge
$AB_2$ chains, (b) substituted polyacetilene, 
with lateral radicals $R^{*}$ as $B^{'}$ sites containing
unpaired residual electrons, and (c) alternate spin chains.}
\label{f1}
\end{center}
\end{figure}

In this section we model and discuss analytical and numerical results 
on the magnetic chains with $AB_2$ and $ABB'$ unit cell topologies
displayed in Figs.~\ref{f1}(a) and (b). 
Eventually, alternate spin chains shown in Fig.~\ref{f1}(c) are also 
considered. 
\newline {\mbox \quad}A rigorous theorem by Lieb~\cite{r29} predicts 
that bipartite $AB_2$ chains modeled through a Hubbard Hamiltonian 
[see Eq.~(\ref{hubbard}) below], with one electron per site on average (half-filled limit) 
and repulsive Coulombian interaction, present average 
ground-state spin per unit cell 
$\hbar/2$ and quantum ferrimagnetic long-range order at $T = 0$~\cite{r4,r29,r15,re16}.
The magnetic excitations on this state have been studied in detail 
both in the weak- and strong-coupling limits~\cite{re17}, and in the 
light of the quantum $AB_2$ Heisenberg model~\cite{re17,carlindo,re18}. 
Further studies have considered the anisotropic~\cite{re19} and 
isotropic~\cite{r5} critical behavior of the quantum $AB_2$ 
Heisenberg model, including its spherical version~\cite{mario}, and the 
statistical mechanics of the classical $AB_2$ Heisenberg model~\cite{carlindo}.
\newline {\mbox \quad}Away from half-filling, 
doped $AB_2$ Hubbard chains were previously studied through 
Hartree-Fock, exact diagonalization and quantum Monte Carlo techniques 
both in the weak- and strong-coupling limits~\cite{r4,rene}, 
including also the $t-J$ model~\cite{re11} using the 
density-matrix renormalization group and recurrent 
variational {\it Ans\"atzes}, and the infinite Coulombian repulsion 
limit~\cite{re12} using exact diagonalization. 
In particular, these chains represent an alternative route to 
reaching $2d$ quantum physics from $1d$ systems \cite{re11,re13}. 

\subsection{Analytical Results}
We start considering the $AB_2$ chain modelled through the one-band Hubbard model, 
which is the simplest lattice model for strongly correlated materials:
\begin{equation}
\mathcal{H}=-t\sum_{ \langle i\alpha,j\beta\rangle,\sigma}c_{i\alpha,\sigma}^{\dagger}c_{j\beta,\sigma}+
U\sum_{i\alpha}n_{i\alpha\uparrow}n_{i\alpha\downarrow},
\label{hubbard}
\end{equation}
where $c_{i\alpha,\sigma}^{\dagger}$ ($c_{i\alpha,\sigma}$) is the creation (annihilation) operator
for electrons with spin $\sigma(=\uparrow,\downarrow$) at site $\alpha=A, B_1 \text{ or } B_2$ 
of the unit cell $i$, 
$t$ is the hopping parameter, $U$ is the intrasite Coulomb repulsion and, 
in the first summation, $i\alpha$ and $j\beta$ are nearest neighbor sites. We define $N$ as the total 
number of sites and $N_c (=N/3)$ as the number of unit cells. We remark that in the 
limit $U=\infty$ \cite{rene}, the Hubbard Hamiltonian reduces to a hopping term with double 
site occupancy excluded.
\par In the tight-binding 
description ($U=0$), this model presents three bands: one flat with $N_c$
localized orbitals with energy $\epsilon=0$ and two dispersive with
$\epsilon_{\pm}=\pm2\sqrt{2}\cos(q/2)$, where $q=2\pi l/N_c$ and $l=0,...,N_c-1$. 
At half-filling ($N_e=N$, where $N_e$ is the number 
of electrons) and $U=0$ the ground state (GS) total spin quantum number $S_g$ is degenerate, with $S_g$ 
ranging from the minimum value (0 or 1/2) to $S_g=|N_B-N_A|/2=N_c/2$, where $N_A$ ($N_B$) 
is the number of sites in the A (B) sublattice. As proved by Lieb \cite{r29}, in the  
general case of any bipartite lattice with $N_B\neq N_A$, the Coulomb repulsion lifts 
this huge degeneracy and selects the state with $S_g=|N_B-N_A|/2$ to be the 
unique GS of the system, apart from the trivial $(2S_g+1)$-fold rotational degeneracy. 
\par
\begin{figure}
\begin{center}
\includegraphics*[width=0.8\linewidth,clip]{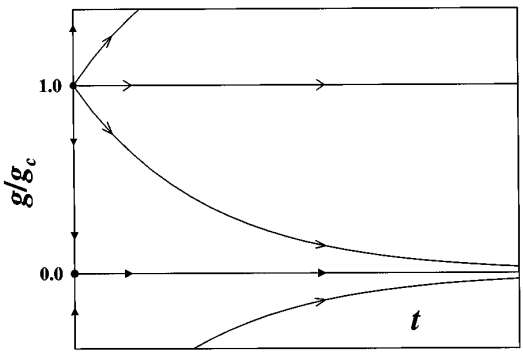}
\caption{Schematic RG ($g,t\propto T$) flux diagram for the $1d$-quantum $z=2$ NL $\sigma$ model.
Semiclassical ($g^*=T^*=0$) and quantum-critical ($g^*=g_c, T^*=0$) fixed points
are shown, as well as the flux lines indicating their stability with respect to infrared perturbations.
The segment $T=0, 0\leq g_0<g_c$, corresponds to the {\it loci} of points in which Lieb's theorem 
is included, with presence os stable ferrimagnetic states of quantum $AB_2$ and alternate spin chains.}
\label{f5} 
\end{center}
\end{figure}

\begin{figure}[b]
\begin{center}
\centerline{\includegraphics*[width=0.8\linewidth,clip]{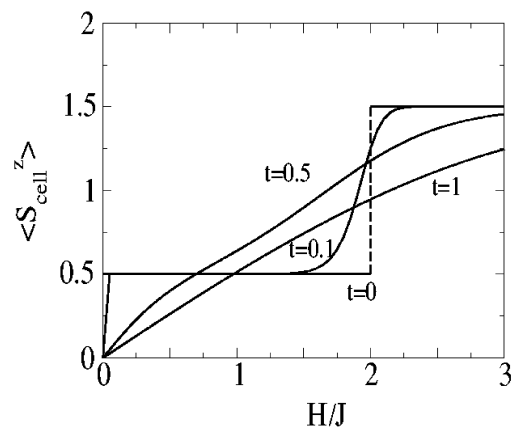}}
\caption{Average spin per unit
cell, $\langle S_{cell}^z \rangle$, in units of $\hbar$, as 
function of the dimensionless magnetic field, $H/J$, and dimensionless
temperature, $t = k_BT /J$, for the ferrimagnetic $S=1/2$ $AB_2$ Ising chain.}
\label{carlindo2c}
\end{center}
\end{figure}

\begin{figure}
\begin{center}
\includegraphics*[width=0.9\linewidth,clip]{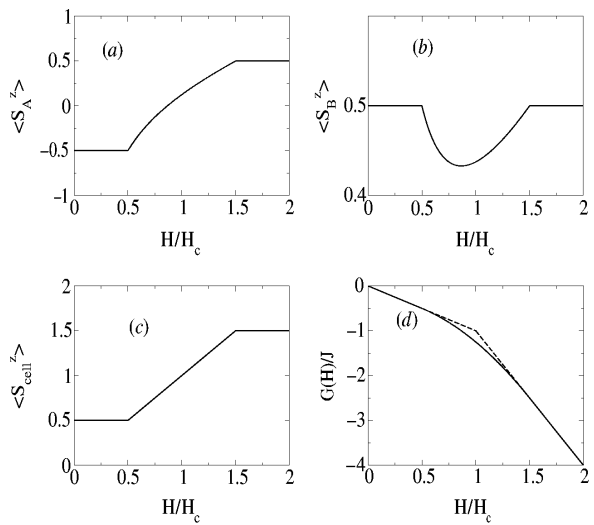}
\caption{Average spin 
at sites A (a), sites B (b) and per unit cell (c), in units of $\hbar$, as       
function of the reduced field, $H/H_c$, for the quantum ferrimagnetic 
$S=1/2$ $AB_2$ Heisenberg chain,
with $H_c = 2J$. (d) The field dependence of the Gibbs free energy shows that the continuous
solution (solid line) for the magnetization is the stable phase. The Ising-like solution is shown for
comparison (dashed line).}
\label{carlindo5} 
\end{center}
\end{figure}

\begin{figure}
\begin{center}
\includegraphics*[width=0.8\linewidth,clip]{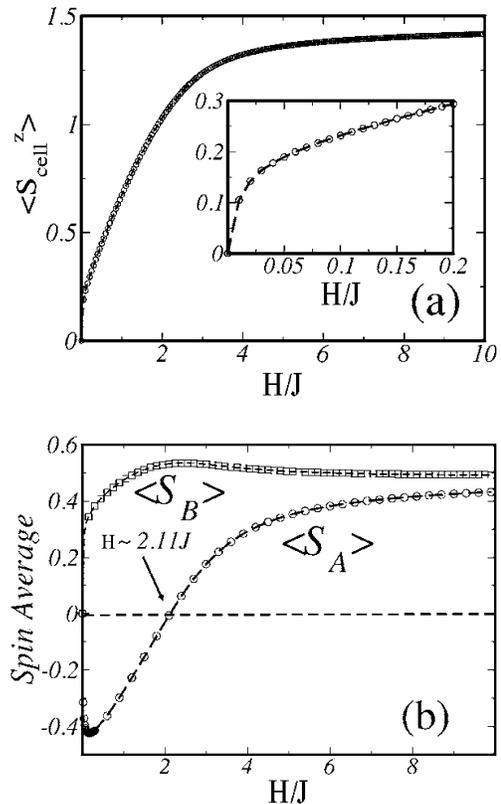}
\caption{(a) Average spin per unit cell, 
$\langle S_{cell}^z \rangle$, and (b) spin averages 
at sites A (-$\bigcirc$-) and B (-$\Box$-), in units
of $\hbar$, as function of $ H / J $, for $g = 0.05J$ and $T = 0.05J$, 
calculated for the quantum spherical $S = 1/2$ $AB_2$ model. 
Inset of (a): very-low-field regime.}
\label{mario3}
\end{center}
\end{figure}

 In the strong coupling limit ($U>>t$) and at half-filling the $AB_2$ 
Hubbard Hamiltonian, Eq.~(\ref{hubbard}), is mapped~\cite{ernestompl,r5} onto 
the quantum $S=1/2$ Heisenberg model with $O(n)$, $n=3$, rotational symmetry:
\begin{equation}
{\cal H} = \sum_{ij}\sum_{\alpha \beta} J_{ij}^{\alpha \beta}   
{\bf S}_{i\alpha}\cdot {\bf S}_{j\beta}~, 
\label{e2}
\label{heisenberg}
\end{equation}
where the localized spins ${\bf S}_{i\alpha}$ interact  
antiferromagnetically through 
$J_{ij}^{\alpha \beta} = J = 4t^2/U > 0$. 
In fact, Lieb and Mattis have shown \cite{physa7} that 
the Heisenberg model in a bipartite lattice has also $S_g=|N_B-N_A|/2$, which 
indicates that $S_g=N_c/2$ for the $AB_2$ Heisenberg model, as in the Hubbard case. 
Eventually, in the presence of an uniform magnetic field ${\bf H}$ along the $z$ direction  
a Zeeman energy term, $- g \mu_B H/\hbar \sum_{i\alpha} S_{i\alpha}^z$, 
is added to Eq.~(\ref{e2}), where $g$ is the gyromagnetic factor and 
$\mu_B$ is the Bohr magneton (in what follows we take units in which 
$g \mu_B \equiv 1$).  
In $H = 0$ the ground-state of the system exhibits~\cite{r29,r4} 
unsaturated FM or ferrimagnetic configurations as indicated in 
Figs.~\ref{f1}(a) and (b), with average spin per unit cell 
$\langle S_{cell}^z \rangle = \hbar/2$ (Lieb's theorem~\cite{r29}). 
In addition, we can also model the alternate 
spin chains shown in Fig.~\ref{f1}(c) by considering in Eq.~(\ref{e2}) 
${\bf S}_{i\alpha} = {\bf S}_{i}$ and ${\bf S}_{i\beta} = {\bf s}_{i}$, 
with $S > s$. Such model systems have been used to describe~\cite{r27,r28} 
a number of organometallic compounds in which, e.g., $S = 5/2$ 
or $S = 2$ and $s = 1/2$. 
\newline {\mbox \quad}The Euclidean action of the partition function 
in a coherent-state ${\bf n}$-field representation~\cite{r30}, 
${\cal Z} = \int {\cal D} {\bf n} \exp (-S_E/\hbar)$, 
is given by $S_E = S_{exc} + S_{WZ} + S_Z$, 
with contributions from exchange and Zeeman interactions, and a 
topological Wess-Zumino term, which is a Berry's phase-like 
term associated with the time evolution of the spin due to 
quantum fluctuations~\cite{r30}.
The low-lying properties of the quantum ferrimagnetic chains 
in Fig.~\ref{f1} are dominated by infrared fluctuations around the 
N\'eel configuration. In order to obtain their effective low-lying action, 
we take staggered dimensionless unit coherent magnetization fields
and split the topological term in a FM and an AF contribution, 
$S_{WZ} = S_{WZ}^{AF} + S_{WZ}^{FM}$, following the spin structure of the 
polymer. Then, taking the continuum limit and integrating 
out the rapidly fluctuating field modes,
we obtain the low-energy effective action, 
$S_{eff} = \int_0^L dx/(2a) \int_0^{\beta \hbar} 
d \tau {\cal L}$, where $L$ is the length of the chain of lattice 
parameter $2a$, and $\beta \equiv (k_BT)^{-1} = it/\hbar$ expresses
the result of a Wick rotation to imaginary times $it$. The Lagrangian density is 
${\cal L} = {\cal L}_{\sigma NL} + {\cal L}_{WZ}^{FM} + {\cal L}_Z + 
{\cal L}_{irrel}$, where 
\begin{equation}
{\cal L}_{\sigma NL} = {\cal L}_{exc} + {\cal L}_{WZ \cdot exc}^{AF} = 
\alpha_x (\partial _x {\bf m} )^2 +  
\alpha_\tau (\partial _\tau {\bf m} )^2~,
\label{e3}
\end{equation}
corresponds to the quantum nonlinear (NL) $\sigma$ model, 
with unit magnetization fields ${\bf m}^2 = 1$,
\begin{equation}
{\cal L}_{WZ}^{FM} = - i {\cal S} \hbar 
\int _0 ^1 d\gamma \partial _\gamma {\bf m}
\cdot ( {\bf m} \times  \partial _\tau {\bf m})~,
\label{e4}
\end{equation}
represents the contribution from the FM Wess-Zumino term, 
and ${\cal L}_{irrel}$ are irrelevant terms in the renormalization group (RG) context. 
In Eqs.~(\ref{e3}) and (\ref{e4}), $\alpha_x = 2JS^2 \hbar ^2 a^2$,
$\alpha_\tau = 1/(8J)$, and ${\cal S} = S$ for the 
$AB_2$ chains, whereas 
$\alpha_x = JS^2 \hbar ^2 a^2$,
$\alpha_\tau = s/(4JS)$, and ${\cal S} = S-s$ for the 
alternate spin chains of Fig.~\ref{f1}. 
The FM Wess-Zumino term is responsible for the ferrimagnetic ground states.
Indeed, ${\cal L}_{WZ}^{FM} = 0$ in either cases of $AB$ or $S = s$ chains.
Those represent usual quantum AF Heisenberg chains, which are known to 
follow Haldane's conjecture~\cite{r31}, i.e., half-integer spin chains
are critical with no long-range order, and integer spin chains are disordered. 
Moreover, the addition of the relevant Wess-Zumino term to the 
quantum NL $\sigma$ model, Eq.~(\ref{e3}), changes its properties 
dramatically since the critical dynamical exponent assumes 
$z = 2$ (nonrelativistic feature), in contrast with the value $z = 1$
found in the relativistic quantum NL $\sigma$ model, 
associated with the $2d$-quantum AF Heisenberg Hamiltonian. 
In fact, Eq.~(\ref{e4}) corresponds to the field-theoretical 
version of the topological constraints imposed by the
polymer structure, as identified by semi-empirical methods~\cite{r1,r13,r14}.
\newline {\mbox \quad}We now perform a momentum-shell low-$T$ 
RG study of the system (see~\cite{r34} and~\cite{r35} 
to similar treatments to the classical and quantum 
$z = 1$ NL $\sigma$ models). First, we 
decompose the magnetization fields into transversal and 
longitudinal components, integrate over the latter one, expand 
the resulting action, $S_{eff} = S^{(2)} + S^{(4)} + ...$, and 
Fourier transform the terms to the momentum-${\bf k}$ and
Matsubara frequency-$\omega_n$ space, with $\omega_n = 2\pi n/u$, 
$n = 0, \pm 1, \pm 2...$, $u = \zeta \beta$, and 
$\zeta = \pi ^2 JS \hbar ^2$ or $\zeta = \pi ^2 J Ss\hbar^2 /(S-s)$
respectively for the $AB_2$ or alternate spin chains. 
The quadratic term in the diagonal field space $\{ \phi ^{\ast}, \phi \}$
reads~\cite{r5}
\begin{widetext}
\begin{equation}
\frac{S^{(2)}({\bf k}, \omega_n)}{\hbar} = 
\sum_{n = - \infty}^{+\infty} \int_{BZ}
\frac{d^d k}{(2\pi)^d} \frac{u}{g_0} 
(k^2 - i \omega_n + hg_0 + 
\frac{\lambda_d \pi^2}{2} \omega_n ^2 - \frac{\rho g_0}{u}) 
\phi ^{\ast}({\bf k}, \omega_n) \phi({\bf k}, \omega_n)~,
\end{equation}
\end{widetext}
where $hg_0 \equiv H \zeta / \hbar$,
the density of degrees of freedom $\rho$ comes from
the integration over the longitudinal components, and the meaning 
of $\lambda_d$ is discussed below. The bare coupling is defined as 
$g_0 \equiv \pi/{\cal S}$ in $d = 1$. The quartic contribution is 
given by~\cite{r5}
\begin{widetext} 
\begin{eqnarray}
\frac{S^{(4)}({\bf k}_i, \omega_{n_i})}{\hbar}& =& 
\sum_{n_1,...,n_4 = - \infty}^{+\infty} \int_{BZ}
[ \prod_{i=1}^4 \frac{d^d k_i}{(2\pi)^d}] \frac{u}{g_0} \\ \nonumber
&&[\frac{1}{2}(-{\bf k}_2 \cdot {\bf k}_4 
+ {\bf k}_2 \cdot {\bf k}_3 
+ {\bf k}_1 \cdot {\bf k}_4 
- {\bf k}_1 \cdot {\bf k}_3) 
+ \frac{i \pi hg_0 }{2^{1/2}} \sum_{j=1}^d 
(k_{j,4} - k_{j,3})  
\\ \nonumber
&&+ \frac{hg_0}{2} - \frac{i}{4}(\omega_{n_3} +  \omega_{n_4}) 
- \frac{\pi}{2^{1/2}} \sum_{j=1}^d ( \omega_{n_3} k_{j,4} - 
\omega_{n_4} k_{j,3})
\\ \nonumber
&&+ \lambda_d \frac{11 \pi}{96 \sqrt{2}} 
(\omega_{n_1}\omega_{n_3} + \omega_{n_1}\omega_{n_4} 
+ \omega_{n_2}\omega_{n_3} + \omega_{n_2}\omega_{n_4}) ]
\\ \nonumber
&&\phi ^{\ast}(1) \phi (2) \phi ^{\ast}(3) \phi (4) 
(2\pi)^d \delta ( {\bf k}_2 + {\bf k}_4 - {\bf k}_1 - {\bf k}_3) 
\delta_{(\omega_{n_2}+\omega_{n_4}),(\omega_{n_1}+\omega_{n_3})}~.
\end{eqnarray}
\end{widetext}
In the sequence, we require that the fluctuation modes 
in the two-point vertex function scales homogeneously 
through a RG scaling transformation, 
$k \rightarrow bk$, $\omega_n \rightarrow b^z\omega_n$, 
with $b \equiv e^\ell$. We also take the fixed point 
$\lambda_d^{\ast} \equiv 0$ for $d = 1$, as a consequence 
of the irrelevance of the $\omega_n^2$-dependent terms in the 
RG context. The one-loop equations for the renormalized coupling $g$, 
and dimensionless temperature $t \equiv g/u$ and magnetic field 
$\bar{h} \equiv hg$ in $d = 1$ and $z = 2$ read~\cite{r5}:
\begin{eqnarray}
\frac{dg}{d\ell} &=& -(d + z - 2)g + 
\frac{\kappa_d}{2} g^2 \coth [g(1+\bar{h})/2t]~,
\nonumber 
\\
\frac{dt}{d\ell} &=& -(d - 2)t + 
\frac{\kappa_d}{2} gt \coth [g(1+\bar{h})/2t]~,
\\
\frac{d\bar{h}}{d\ell} &=& 2\bar{h}~, \quad \frac{du}{d\ell} = -zu~,
\quad u = \frac{g}{t}~.
\nonumber
\end{eqnarray} 
We thus obtain the semiclassical fixed point: $g^{\ast}= 
t^{\ast}= \bar{h}^{\ast}= 0$, since $g^{\ast}= 0$ implies in 
${\cal S} \rightarrow \infty$, and the quantum critical fixed point~\cite{r35}:
$g^{\ast} \equiv g_c = 2 \pi,  
t^{\ast}= \bar{h}^{\ast}= 0$. The former describes a $1d$-classical 
Heisenberg ferromagnet with quantum corrections, whereas the 
latter is identified with a classical Heisenberg model in $d + z = 3$ dimensions. 
The analysis of stability shows that both fixed points are unstable 
under thermal fluctuations, but only the semiclassical fixed point 
is stable under infrared quantum fluctuations, as shown in Fig.~\ref{f5}. 
\newline {\mbox \quad}By studying the correlation length $\xi$ 
and magnetic susceptibility $\chi$ we identify~\cite{r5} three 
distinct quantum regimes. 
As $T \rightarrow 0$ and $g > g_c$, the 
quantum $z = 2$ NL $\sigma$ model is in a quantum
disordered phase, whereas for $g < g_c$ its ground state has long-range order, with both 
quantum and thermal fluctuations playing important roles. 
For $g = g_c$ and $T \rightarrow 0$ the system approaches the quantum critical fixed point 
characterized by the extinction of the spin-wave modes and the absence of 
long-range order. As shown in Fig.~\ref{f5}, the quantum critical region is defined 
by the crossover lines $T \sim |g-g_c|^\phi$, where $\phi = z \nu_3$, 
with $\nu_3$ the $3d$-Heisenberg correlation-length exponent. 
In particular, we find~\cite{r5} the following low-$T$ behavior 
in the quantum critical region:
\begin{equation}
\xi \sim T^{-1/z}~, \quad \chi \sim T^{-1}~, \quad C \sim T^{d/z}~,
\label{e9}
\end{equation}
with the standard result for the low-$T$ quantum critical specific heat $C$ 
and exponents satisfying scaling relations proper of this region~\cite{r33,r36}. 
Similarly, for the $T \rightarrow 0$, $g < g_c$ semiclassical region, we find
\begin{equation}
\xi \sim T^{-1}~, \quad \chi \sim T^{-2}~, \quad C \sim T^{1/2}~,
\label{e10}
\end{equation}
where the specific heat is determined by the spin wave contribution. 
In Eqs.~(\ref{e9}) and (\ref{e10}) the amplitudes of the observables 
cannot be completely fixed by the RG procedure. 
We thus identify the asymptotic low-$T$ critical behavior 
described by Eq.~(\ref{e10})
with that of the quantum $S=1/2$ $AB_2$ 
and alternate spin ferrimagnetic chains, as well as that of the $1d$-quantum 
$S=1/2$ Heisenberg ferromagnet, such as the organic ferromagnetic 
compound $p$-NPNN~\cite{r11}. 
In addition, an interesting question arises 
regarding the access of such $AB_2$ chains in the 
half-filled strong-coupling limit to the 
quantum disordered and quantum critical regimes of the quantum $z = 2$
NL $\sigma$ model. This scenario, if accomplished,  
might involve the presence of extra frustrated couplings in the 
unit cell structure. 
In any case, we would like to mention that 
our predicted one-loop critical behaviors for the renormalized classical and 
quantum critical fixed points are in agreement with 
those of the FM transition in $1d$ itinerant electron systems 
in the context of a Luttinger liquid framework~\cite{kyang}.
However, while our localized spin disordered phase is gapped, 
the quantum disordered phase in Ref.~\cite{kyang} behaves as an 
ordinary gapless Luttinger liquid. Obviously, further theoretical work 
is needed in order to clarify the physical scenario predicted 
for the critical behavior of the quantum NL $\sigma$ model with a FM Wess-Zumino term 
due to the $AB_2$ topology.
\newline {\mbox \quad}In order to improve the understanding 
of the role of the quantum and 
thermal fluctuations, topology, and spin symmetry to the properties 
of the ferrimagnetic $AB_2$ chains, we have also performed a number 
of analytical studies using Ising, Heisenberg and spherical Hamiltonians 
as model systems in this unit cell structure~\cite{carlindo,fred,mario}. 
\newline {\mbox \quad}First, by regarding the 
spin operators in Eq.~(\ref{e2}) as Ising variables, $S_{i\alpha} = 
\pm \hbar/2$, we apply~\cite{carlindo} the RG decimation of $B$ sites and obtain 
the exact Gibbs free energy as function of the effective coupling $J^{\ast}$ 
and field $H^{\ast}$. At zero field the ground state result $J^{\ast} = -2J < 0$ 
for the effective coupling between $A$ sites 
indicates the presence of a ferrimagnetic structure with 
average spins at $B$ sites 
pointing opposite to those at $A$ sites, 
implying in a unit cell average spin $\langle S_{cell}^z \rangle = \hbar/2$. 
As $H$ increases at $T = 0$, we notice that $J^{\ast}$ increases linearly 
with $H$ and vanishes for $H \ge J$; 
conversely,  $H^{\ast}$ first decreases linearly with $H$ for $H < J$, and 
then increases also linearly, changing sign at the critical field $H_c = 2J$.
At $H = H_c$ a first order transition occurs with 
a discontinuous change of $\langle S_{cell}^z \rangle$ 
from $\hbar/2$ to its saturated 
value $3\hbar/2$ 
(see Fig.~\ref{carlindo2c}). At finite temperatures the described effects
are less dramatic, and the unit cell average spin grows continuously  
with the field from 0 at $H= 0$ (disordered state at finite $T$) 
to the saturated value $3\hbar/2$ as $H \rightarrow \infty$.  
The $H = 0$ results are corroborated by the calculation of the 
two-spin correlation function, which is related to the susceptibility 
through the fluctuation-dissipation theorem. Actually, we have also found 
that $\chi \sim \xi \sim \exp [J/(k_BT)] / (k_BT)$, leading to the 
relation between the corresponding critical exponents $\gamma = \nu = 2 - \alpha$, 
and from the behavior of the correlation function at $T = 0$ and 
magnetization at $T = 0$ and $H \rightarrow 0$  
it follows that $\eta = 1$ and $\delta = \infty$. 
This set of exponents belongs to the same class of universality 
of decorated $1d$ ferromagnetic Ising systems~\cite{carlindo22}.
\newline {\mbox \quad}Now, by considering Heisenberg spins in a classical context, 
in which quantum fluctuations are absent, the low-$T$ and high-$T$ limits 
of the $H = 0$ free energy and two-spin correlation functions have been  
calculated~\cite{carlindo}. 
It is instructive to compare the $T \rightarrow 0$ result obtained for the 
classical ferrimagnetic $AB_2$ Heisenberg chain,
\begin{equation}
\chi = \frac{2}{9} \frac{J}{(k_BT)^2}[ 1 - \frac{13}{4} \frac{k_BT}{J} + ...]~,
\end{equation}
with that of the classical ferromagnetic linear Heisenberg chain~\cite{carlindo19},
\begin{equation}
\chi = \frac{2}{3} \frac{J}{(k_BT)^2}[ 1 - \frac{1}{2} \frac{k_BT}{J} + ...]~,
\end{equation}
and that of Takahashi~\cite{r11} for the quantum ferromagnetic spin-1/2 linear 
Heisenberg chain,
\begin{equation}
\chi = \frac{2}{3} \frac{J}{(k_BT)^2}[ 1 
- \frac{3\zeta(1/2)}{(2\pi)^{1/2}} \frac{(k_BT)^{1/2}}{J^{1/2}} + 
\frac{3\zeta^2(1/2)}{2\pi} \frac{k_BT}{J} + ...]~, \label{enovo}
\end{equation}
where $\zeta(1/2)/(2\pi)^{1/2} \approx -0.583$.
At low-$T$ Fisher's and Takahashi's leading terms coincide 
and are three times larger than that of the $AB_2$ chain, due to 
the unit-cell topology effect. Moreover, the second term in Eq.~(\ref{enovo}), 
absent in the classical models, is related to the fixing of the  
anomalous entropy and specific heat classical behaviors  
when quantum fluctuations are not present.  
We notice that this $\chi \sim T^{-2}$ 
leading result as $T \rightarrow 0$ has  
also been obtained for the semiclassical fixed point 
of the quantum $z = 2$ NL $\sigma$ model, related to the 
quantum $AB_2$ Heisenberg chains (see above). 
\newline {\mbox \quad}By taking quantum fluctuations into account, 
we also calculate~\cite{fred} the three spin-wave modes of 
the quantum spin-1/2 $AB_2$ Heisenberg model 
using the Holstein-Primmakov 
transformation and subsequent diagonalization via 
the Bogoliubov-Tyablikov method, namely, one non-dispersive optical, $\epsilon_k^{a}$,  
one dispersive optical, $\epsilon_k^{b}$, and one acoustical mode, $\epsilon_k^{c}$:
\begin{eqnarray}
\epsilon_{k}^{a} &=& J + H~, \\ \nonumber 
\epsilon_{k}^{b,c} &=& \frac{J}{2} \{ \pm 1 + [1 + 8 \sin ^2 (ka)]^{1/2} \} \mp H~;
\label{swmodes}
\end{eqnarray}
notice in the acoustical mode the presence of a quadratic ferromagnetic dispersion relation 
$\epsilon_k^c = 2J(ka)^2$, 
$ka \ll 1$, $H = 0$. From this result, corrections due to 
quantum fluctuations to the average values $\langle S_{B}^z 
\rangle = - \langle S_{A}^z \rangle = \hbar/2$ are derived, 
although  
the result of Lieb's theorem, $\langle S_{cell} \rangle = \hbar/2$, remains true.     
In a mean-field 
approach~\cite{carlindo,fred}, we relate 
the quantum thermal spin averages at sites $A$ and 
$B$ to the respective Weiss molecular fields and find, in the simplest 
case in which they are assumed to be parallel to the $z$ direction (Ising-like 
solution): 
$\langle S_B^z \rangle = \hbar/2$ for all $H \ge 0$, and 
$\langle S_A^z \rangle = -\hbar/2$ for $0 \le H < H_c$, whereas 
$\langle S_A^z \rangle = \hbar/2$ for $H > H_c$. 
We notice that $H_c = 2J$ is the critical field below which 
the ferrimagnetic ordering is favoured, in agreement with the above result 
for the spin-1/2 $AB_2$ Ising model. Indeed, the unit cell spin reads  
$\langle S_{cell}^z \rangle = \hbar/2$ for $0 < H < H_c$, and  
$\langle S_{cell}^z \rangle = 3\hbar/2$ for $H > H_c$. 
On the other hand, in the case the $x$ and $y$ components are also considered, 
a quite interesting scenario emerges, with 
$\langle S_{\alpha}^z \rangle = \mp \hbar/2$ for $0 \le H < H_c/2$, and  
$\langle S_{\alpha}^z \rangle = \hbar/2$ for $H > 3H_c/2$, 
where the plus (minus) sign refers to $\alpha = B$ ($A$) sites; 
for intermediate fields, $H_c/2 \le H \le 3H_c/2$, one has  
that $\langle S_{A}^z \rangle = - \hbar [3 H_c/(4H) - 
H/H_c]/2$ and $\langle S_{B}^z \rangle = \hbar [3 H_c/(8H) + 
H/(2H_c)]/2$. These results imply in the unit cell average spin 
$\hbar/2$ for $H < H_c/2$, with ferrimagnetism sustained, 
and the saturated $3\hbar/2$ value for $H > 3H_c/2$, as in the 
Ising-like solution. A 
linear increase with $H$ arises for intermediate fields: 
$\langle S_{cell}^z \rangle = H/H_c$, for $H_c/2 < H < 3H_c/2$ 
(see Fig.~\ref{carlindo5}).  
In this regime the average spin at sites $A$ continuously rotates 
seeking a full alignment with $H$, accompanied by a rotation of the 
spins at sites $B$, such that the transversal spin components 
at sites $A$ and $B$ always cancel out. To achieve this cancellation 
the spins at sites $B$ rotate in the opposite direction up to 
a maximum polar angle $\theta = \pi/6$ and then rotate back 
[see Fig.~\ref{carlindo5}(b)]. These results are corroborated by 
the analysis of the Gibbs free energy [Fig.~\ref{carlindo5}(d)]. 
\newline {\mbox \quad}At last, a spherical version of the quantum 
$AB_2$ spin Hamiltonian has also been studied~\cite{mario}. 
For this purpose, a chemical potential ($\mu$) term is added to Eq.~(\ref{e2}) 
in order to take care of the spherical constraint, $\sum_{i\alpha} \langle 
S_{i\alpha}^2 \rangle = N/4$, where $N$ is the total number of sites. 
Quantum fluctuations are introduced, associated with a quantum coupling 
parameter $g$, through a kinetic energy term $(g/2) \sum_{i\alpha} P_{i\alpha}^2$, 
in which $P_{i\alpha}$ are momentum operators canonically conjugated 
to each spin degree of freedom. By diagonalizing the Hamiltonian 
in a space of proper bosonic operators, we obtain two dispersive eigenmodes, also 
present in the linear AF spherical model, and a flatband induced 
by the $AB_2$ topology. At the only critical point, $g = T = H = 0$, 
the ferrimagnetic long-range order is present with 
$\langle S_{B} \rangle = -\langle S_{A} \rangle/\sqrt{2} = \hbar\sqrt{3}/4$
and $\langle S_{cell}^z \rangle = \hbar(\sqrt{2} - 1)\sqrt{3/8}$. 
Interestingly, in the quantum $AB_2$ spherical case 
the average spin per unit cell is less than $\hbar/2$, 
in contrast with the result of Lieb's theorem for the $AB_2$ Hubbard model 
in the strong-coupling half-filled limit and the quantum $AB_2$ Heisenberg 
chain with AF couplings. Calculation of the correlation functions at $g = T = H = 0$ 
show that they are distance independent and finite, consistently 
with the ferrimagnetic order and the spherical constraint. 
Outside this critical point, for any 
finite $g$, $T$ or $H$, quantum and/or thermal fluctuations destroy 
the long-range order in the system, which in this case displays    
a finite maximum in the susceptibility. In this regime spins remain ferrimagnetically 
short-range ordered to some extent in the $\{ g, T, H \}$ parameter 
space as a consequence of the AF interaction and the $AB_2$ topology. 
Indeed, we notice in Fig.~\ref{mario3} that, although the field-induced 
unit cell average spin, $\langle S_{cell}^z \rangle$, 
displays quantum paramagnetic behavior for any finite $g$ or $T$, 
the spins at sites $A$ and $B$ can display opposite orientations 
depending on the values of $g$, $T$ and $H$. Therefore, for special 
regions of the parameter space $\{ g, T, H \}$ spins at sites $A$ 
points antiparallel with respect to those at sites $B$, thus giving 
rise to a rapid 
increase in the unit cell average spin for very low $H$
and a field-induced short-range ferrimagnetism, which is destroyed 
for large $g$, $T$ or $H$. 
In addition, to better characterize the approach to 
the $g = T = H = 0$ critical point, we have considered several paths.
For $T \rightarrow 0$ and $g = H = 0$ the susceptibility behaves as 
$\chi \sim T^{-2}$, as also found in several classical and quantum 
spherical and Heisenberg models (see above). On the other hand, 
for $H \rightarrow 0$ and $g = T = 0$, we find $\chi \sim H^{-1}$, 
and for $g \rightarrow 0$ and $T = H = 0$, $\chi \sim 
\exp (cg^{-1/2})$,
where $c$ is a constant, evidencing an essential singularity due to quantum fluctuations. In any
path, the relation $\chi \sim \xi^2$ is satisfied. We also mention that 
the known drawback  
of classical spherical models as $T \rightarrow 0$  
regarding the third law of thermodynamics (finite specific heat and
diverging entropy) is fixed in the presence of quantum 
fluctuations, $g \not = 0$.   
\subsection{Numerical Results}
\par
\begin{figure*}
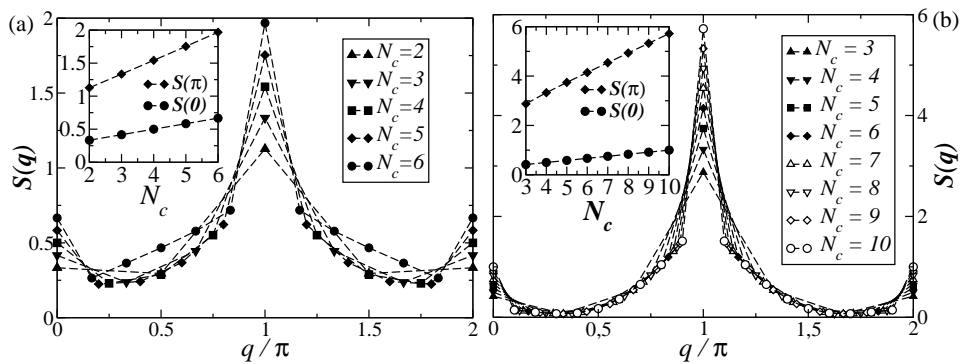

\begin{center}
\includegraphics[width=0.35\linewidth,clip]{Figphys2a.eps}
\includegraphics[width=0.35\linewidth,clip]{Figphys1a.eps}
\caption{Magnetic structure factor $S(q)$ for (a) the $AB_{2}$ Hubbard chain
with $U=2t$ and (b) the $AB_2$ Heisenberg chain. The inset
presents the size dependence of the ferromagnetic [$S(0)$] and antiferromagnetic
[$S(\pi)$] peaks. Dashed lines are guides for the eye.}
\label{figphys2a}
\end{center}
\end{figure*}

 The ferrimagnetic ordering can be probed through the magnetic structure factor:
\begin{equation}
S(q)=\frac{1}{N}\sum_{i,j} e^{iq(x_{i}-x_{j})} \langle \mathbf{S}_{i}\cdot\mathbf{S}_{j}\rangle~.
\end{equation}
The condition for a long-range ferromagnetic ordering is that $S(0)\sim N$;
while $S(\pi)\sim N$ in a long-range antiferromagnetically ordered state;
A {\it ferrimagnetic} long-range ordering fullfill these two conditions:
$S(0)\sim N$ and $S(\pi)\sim N$. 
This is the case for the $AB_2$ Hubbard and Heisenberg chains 
 as exemplified in Fig. \ref{figphys2a}
through the exact diagonalization (ED) of finite clusters.
Due to the critical nature of both chains at low temperatures,
the correlation length $\xi$ and $\chi(q=0)=S(q=0)/(k_B T)$ satisfy power law behavior:
$\xi\sim T^{-\nu}$ and $\chi\sim T^{-\gamma}$ as $T\rightarrow 0$.
Since $\xi \sim N$ at $T=0$, using scaling arguments and the results of Fig. \ref{figphys2a},
we have $T^{-\gamma}\sim T^{-\nu}/T$, i. e., $\gamma-\nu=1$, in agreement with
the values $\gamma=2$ and $\nu=1$ derived using renormalization group
techniques. Furthermore, the ferrimagnetism of the model was also manifested 
through Hartree-Fock and Quantum Monte Carlo methods \cite{r4}.

\par
\begin{figure}[b]
\begin{center}
\centerline{\includegraphics[width=0.75\linewidth,clip]{Figphys4a.eps}}
\centerline{\includegraphics[width=0.75\linewidth,clip]{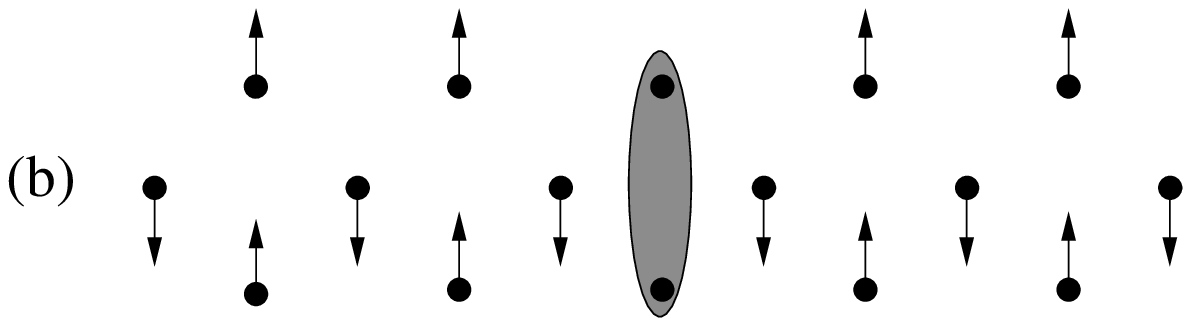}}
\caption{(a) Ferromagnetic (F) and antiferromagnetic (AF) spin wave modes of the Heisenberg 
$AB_{2}$ chain for $N_c=10$ (circles), 8 (triangles down), 6 (triangles up). Solid lines
are the Linear Spin Wave results from Ref. \cite{fred};
dashed lines are guides to the eye. (b) Illustration of the F2 mode: ellipse indicates 
a localized singlet state.}
\label{figphys4a}
\end{center}
\end{figure}

 Systems with a ferrimagnetic GS naturally have ferromagnetic (lowering the GS spin) and
antiferromagnetic (rising the GS spin)
 magnons as their elementary magnetic excitations.
 The $AB_2$ chain have three spin wave branches \cite{re17}:
 an antiferromagnetic mode (AF mode), defined as
 $\Delta E_{S+}(q)=E(S^{z}=S_{g}+1,q)-E_{GS}$;
 and two ferromagnetic ones (F1 and F2 modes), derived from
 $\Delta E_{S-}(q)=E(S^{z}=S_{g}-1,q)-E_{GS}$,
 where $E_{GS}$ is the GS energy and $E(S^{z},q)$
 are lowest energies in the sector $\{S^z,q\}$,
 with the lattice wave-vector $q=2\pi l/N_c$, where $l=0,1,...,N_c-1$. 
 These modes are depicted in Fig. \ref{figphys4a} (a) for the Heisenberg model:
 the AF mode has a gap $\Delta_{S+}=1.7591J$;
 the gapless F1 mode is the Goldstone mode,
 consistent with the symmetry broken
 phase of the chain; and the F2 mode has a gap $\Delta_{S-}=1.0004J$.
 The gapped F2 branch is flat and is associated with the formation of a singlet state
 between the B sites in one cell, while the other cells have B sites in triplet states, 
 as illustrated in Fig. \ref{figphys4a} (b). The localized nature of the excitation 
 is associated with the Hamiltonian invariance under the exchange of the B 
 sites of any cell, this symmetry implies that the many-body wave function has a definite 
 parity under the exchange of the spacial variables associated with these sites.
 Since these dispersive modes preserve the local triplet
bond, they are identical to those found in the
spin-$\frac{1}{2}$/spin-$1$ chain \cite{spin1spin2}.
Surprisingly, Linear Spin Wave Theory \cite{fred} predicts
that $\Delta_{S-}=1$,
very close to our estimated value: $\Delta_{S-}=1.0004J$.
Moreover, a good agreement is found for the gapless F1 branch
in the long wave-length limit.
However,
both LSWT and mean field theory \cite{carlindo} predicts $\Delta_{S+}=1$,
deviating from our estimated exact diagonalization value: $\Delta_{S+}=1.7591J$,
which is in excellent agreement with
numerical and analytical calculations for the
spin-$\frac{1}{2}$/spin-$1$ chain \cite{spin1spin2}.
On the other hand, the
Interacting Spin Wave Theory \cite{re18}
derives a better result for $\Delta_{S+}$,
but it implies in a higher shift for $\Delta_{S-}$ (flat mode)
not observed in our data of Fig. \ref{figphys4a} (a).

\begin{figure}
\begin{center}
\centerline{\includegraphics*[width=0.8\linewidth]{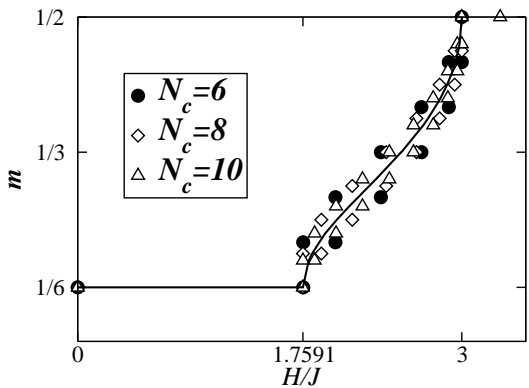}}
\caption{Magnetization as a function of the dimensionless applied magnetic field $H/J$ for
the $AB_{2}$ Heisenberg chain. The full line is a curve traced 
from the midpoints of the steps found in 
the finite size results, except at plateau regions.}
\label{figphysa5a}
\end{center}
\end{figure}

On the other hand, the AF mode is relevant in the analysis of the response
to an applied magnetic field $\textbf{H}$. 
The AF gap found above is responsible for
a plateau in the curve of the magnetization per spin [$m(H)= \langle S^z \rangle/(N\hbar)$] as
a function of $H$.
In fact, it has been shown \cite{physa18}
that if $\nu(s-m)=\mbox{integer},$
a plateau may appear in the magnetization curve of the Heisenberg model.
In the last equation, $s$ is the site spin quantum number and $\nu$ 
is the number of sites in one unit
cell of the GS for a given value of $H$. The $AB_2$ Heisenberg chain
has $s=1/2$ and three sites per unit cell ($\nu=3$);
so, unless the system spontaneously breaks the translation
invariance, we expect
plateaus at $m=1/6$ and $m=1/2$. 
This is indeed what is observed in Fig. \ref{figphysa5a}.
The plateau width at $m=1/6$ is exactly given by $\Delta_{S+}$, 
and is a measure of the
stability of the ferrimagnetic phase. 
For higher fields, the magnetization
increases in the expected way \cite{physa18}, as shown by the full line 
in Fig. \ref{figphysa5a}, before saturation
at $m=1/2$ for $H=3J$. This field-dependent 
behavior contrasts with the linear one predicted by 
mean-field theory \cite{carlindo} shown in Fig. \ref{carlindo2c}. Exact diagonalization results \cite{re17}
indicate that the antiferromagnetic spin gap 
and, consequently, the plateau at $m=1/6$ exists for any finite 
value of $U$, with the plateau width ($\Delta_{S+}$) 
nullifying as $U^2$ in the limit $U\rightarrow 0$.  

\begin{figure}[t]
\begin{center}
\includegraphics*[width=1.0\linewidth]{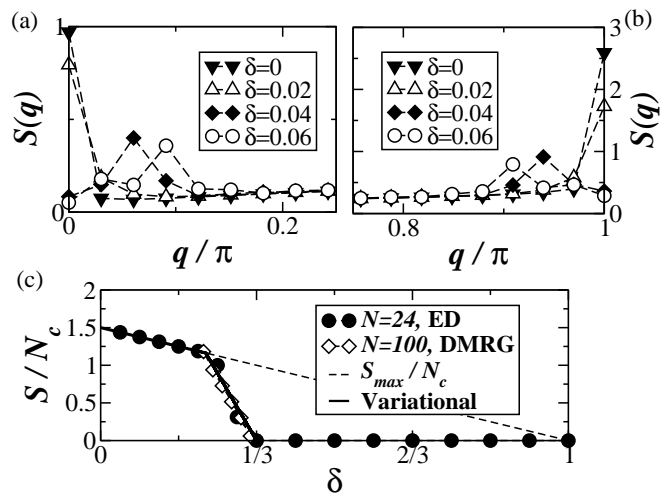}
\caption{ (a) and (b): Magnetic structure factor for $U=2t$ and $N=100$, using DMRG, in the underdoped 
region. (c) Total spin per cell $S_g/N_c$ as function of $\delta$ for $U=\infty$; the variational 
approach is described in Ref. \cite{rene}}
\label{figrene3}
\end{center}
\end{figure}

Away from half-filling \cite{r4,rene}, the $AB_2$ Hubbard model 
exhibits a rich phase diagram depending on the electronic density, 
$n\equiv N_e/N$, or doping $\delta(=1-n)$ from half-filling, 
and the Coulomb coupling $U$. The doped region was analyzed 
through Hartree-Fock \cite{r4}, exact diagonalization \cite{r4,rene} 
and density matrix renormalization 
group (DMRG) \cite{rene}, which is the state-of-the-art method \cite{re34}
for the study of the GS properties of one-dimensional quantum lattice models.
DMRG results suggest that in the underdoped region and for $U=2t$, the 
ferrimagnetic phase sustains up to $\delta\sim 0.02$ while for 
$0.02\lesssim \delta \lesssim 0.07$ hole itinerancy  
promotes incommensurate spin correlations (a spiral phase \cite{spiral}) with a
$\delta$-dependent peak position in the magnetic structure factor, 
as shown in Fig. \ref{figrene3} (a) and (b). 
For $U=\infty$ and $\delta=0$ the GS total spin is degenerate,
 whereas for $0<\delta \lesssim 0.225$ hole itinerancy (Nagaoka mechanism \cite{nagaoka,tasaki}) 
 sets a fully polarized GS, as shown in Fig. \ref{figrene3} (c). 

\begin{figure}
\begin{center}
\centerline{\includegraphics[width=0.9\linewidth,clip]{Figrene5.eps}}
\centerline{\includegraphics[width=0.9\linewidth,clip]{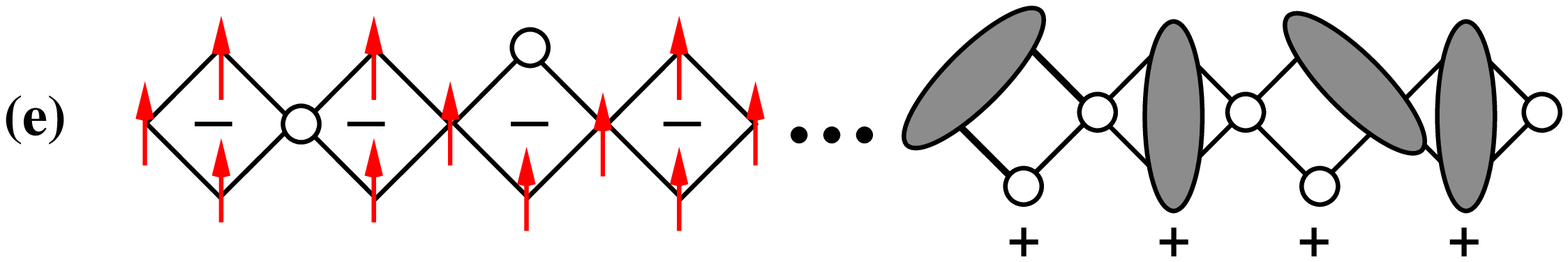}}
\centerline{\includegraphics[width=0.8\linewidth,clip]{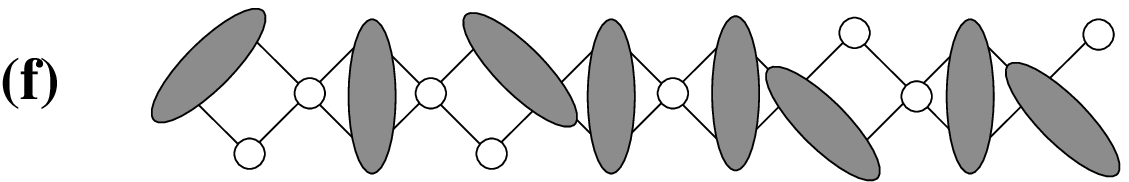}}
\caption{GS properties at $\delta=0.18$ ($U=2t$) and $\delta=0.28$ ($U=\infty$) 
for $N=100$ using DMRG. (a) Spin correlation function $\langle 
\mathbf{S}_{1}\cdot\mathbf{S}_{i}\rangle$ for $U=2t$. (b) Expectation value of 
$S^z_i$ for $U=\infty$ in the sector $S^z=S_g$. Spin correlation function 
$\langle \mathbf{S}_{B_1}\cdot\mathbf{S}_{B_2}\rangle_i$ for (c) $U=2t$ and (d) 
$U=\infty$. $-$($+$) indicates odd (even) local parity. Dashed lines are guides 
to the eye. Illustration of the GS for (e) $U=\infty$ in the phase-separated regime 
and (f) at $\delta=1/3$: singlet bonds are represented by ellipses and holes by circles.}
\label{figphaseseparation}
\end{center}
\end{figure}

\begin{figure}
\begin{center}
\centerline{\includegraphics[width=0.9\linewidth,clip]{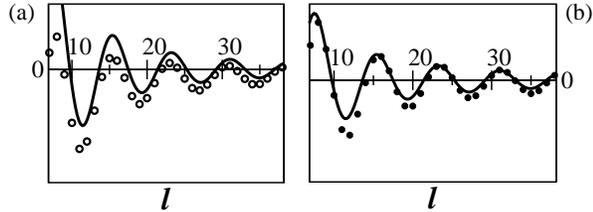}}
\caption{Spin correlation functions $C(l)$ for (a) $U=2t$ and (b) $U=\infty$ at 
$\delta=88/106$ for $N=106$ using DMRG: solid lines are fittings using Eq. (\ref{cll}).}
\label{figrene13}
\end{center}
\end{figure}

For higher doping, the system phase separates \cite{sepfases} into coexisting metallic and insulating
 phases for $\delta_{PS}(U)\lesssim \delta <1/3$ [with $\delta_{PS}(\infty)\approx 0.225$ and 
 $\delta_{PS}(2)\approx 0.07$].
In fact, the Hartree-Fock 
solution is unstable in this region and a Maxwell construction is needed \cite{r4}. 
The local parity symmetry is even (odd) in the insulating (metallic) phase.
In Fig. \ref{figphaseseparation} we present spin correlation functions 
at $\delta=0.18$ ($\delta=0.28$) and $U=2$ ($U=\infty$) calculated through DMRG (for these parameter values, the 
system is found in a phase separated state).
In Fig. \ref{figphaseseparation}(a) the correlation function between 
the spin at the extrem site of the phase with odd parities and the 
others spins, $\langle \mathbf{S}_{1}\cdot\mathbf{S}_{i} \rangle$, 
evidences the spiral phase for $U=2t$. For $U=\infty$ the local magnetization
at sites A and $B_1+B_2$ shown in Fig. \ref{figphaseseparation}(b) 
clearly displays the coexistence between the Nagaoka ferromagnetic phase 
and a paramagnetic one. Notice that, in the Nagaoka phase, 
the magnetization displays spatially modulated profiles due to hole itinerancy.   
Further, the local parity symmetries of the two coexisting phases are 
manifested in the correlation function between B spins 
at the same cell, $\langle \mathbf{S}_{B_1}\cdot\mathbf{S}_{B_2}\rangle$,
shown in Figs. \ref{figphaseseparation}(c) and (d): in the phase with odd (even) parities 
cell triplet (singlet) states predominate. 
In Fig. \ref{figphaseseparation}(e) we illustrate the phase separation for $U=\infty$.
 DMRG and exact diagonalization results \cite{rene} indicate
that the phase separation region ends precisely at $\delta=1/3$. For this 
doping the electronic system presents finite spin and charge gaps with very short ranged 
correlations and is well described by a short-ranged resonating-valence bond (RVB) state \cite{re44},
with the electrons correlated basically within a cell,
as illustrated in Fig. \ref{figphaseseparation}(f). A crossover 
region is observed for $1/3\leq\delta\leq 2/3$, while a Luttinger-liquid behavior \cite{re9} 
can be explicitly characterized for $\delta>2/3$. Luttinger liquids 
are paramagnetic metals in one-dimension exhibiting power-law decay of the 
charge and spin correlation functions and the separation of the 
charge and spin excitation modes. In particular, the asymptotic 
behavior of the spin correlation function is given by
\begin{equation}
C_{LL}(l)\sim \frac{\cos(2k_F l)[\ln(l)]^{1/2}}{l^{1+K_\rho}},
\label{cll}
\end{equation}
where $k_F$ is the Fermi wave vector and $K_\rho$ is the model-dependent exponent. 
The predicted behavior in Eq. (\ref{cll}) fits very well the DMRG data 
at $\delta=88/106$ using $K_\rho=0.89$ ($U=2t$) and $K_\rho=0.57$ ($U=\infty$), as shown in 
Fig. \ref{figrene13}.
We remark that these values are close to 1 (noninteracting fermions) 
and 1/2 (noninteracting spinless fermions) for $U=2t$ and $U=\infty$, respectively. 

In addition, we mention 
that the commensurate doping $\delta=2/3$ is insulating,  
with a charge gap nullifying with $U$ in a similar manner as the one of the 
 Hubbard model in a linear chain at half-filling \cite{re47}, while 
 the spin excitation is gapless.
\section{Conclusions}
In this contribution on the celebration of the 80th birthday anniversary 
of Prof. Ricardo Ferreira, we have presented a brief 
review of the main experimental and theoretical 
achievements on quasi-one-dimensional 
magnetic compounds, featuring those with $AB_2$ unit cell structure. 
\newline {\mbox \quad}As reported, this has been an area of intense activity,
particularly since the first experimental announcements in the 
mid 80's~\cite{r1,r10,r22,yo1}. Nowadays several groups all 
over the world, involving chemists, physicists, and material scientists, 
are engaged in the characterization and description of properties 
of the already known materials, as well as doing great efforts 
towards the design and synthesis of new compounds, with 
novel properties suitable for technological applications~\cite{r37}. 
\newline {\mbox \quad}From the fundamental point of view, 
these compounds have been used as a laboratory in 
which many theoretical concepts and predictions in the 
field of low-dimensional materials have been tested. 
\newline {\mbox \quad}In conclusion, it seems clear that this 
interdisciplinary research area will remain an exciting and topical one 
for many years to come, offering new challenges both from the 
scientific and technological aspects.  
\section{Acknowledgments}
We thank A. M. S. Mac\^edo, M. C. dos Santos,  C. A. Mac\^edo and 
F.~B. de Brito 
for collaboration in several stages of this research. 
Work supported by CNPq, CAPES, Finep and FACEPE 
(Brazilian agencies).
\bibliography{ref}
\end{document}